# Thinging for Computational Thinking

Sabah Al-Fedaghi[1], Ali Abdullah Alkhaldi[2]
Computer Engineering Department
Kuwait University, Kuwait

*Abstract*—This paper examines conceptual models and their application to computational thinking. Computational thinking is a fundamental skill for everybody, not just for computer scientists. It has been promoted as skills that are as fundamental for all as numeracy and literacy. According to authorities in the field, the best way to characterize computational thinking is the way in which computer scientists think and the manner in which they reason how computer scientists think for the rest of us. Core concepts in computational thinking include such notions as algorithmic thinking, abstraction, decomposition, and generalization. This raises several issues and challenges that still need to be addressed, including the fundamental characteristics of computational thinking and its relationship with modeling patterns (e.g., object-oriented) that lead to programming/coding. Thinking pattern refers to recurring templates used by designers in thinking. In this paper, we propose a representation of thinking activity by adopting a thinking pattern called *thinging* that utilizes a diagrammatic technique called thinging machine (TM). We claim that thinging is a valuable process as a fundamental skill for everybody in computational thinking. The viability of such a proclamation is illustrated through examples and a case study.

## I. Introduction

The cognitive faculty of thinking [1] involves processes by which we reason and solve problems. "Computational thinking is a fundamental skill for everybody, not just for computer scientists. To reading, writing, and arithmetic, we should add computational thinking to every child's analytic ability" [2]. Computational thinking is distanced from digital literacy/competence, as it focuses on problem-solving processes and methods and on creating computable solutions [3]. It has been promoted as skills that are as "fundamental for all as numeracy and literacy" [3]. It goes beyond introductory knowledge of computing to treat computer science as an essential part of education today and presents a distinct form of thought, separate from these other academic disciplines, where diagrammatic techniques are used in analysis and strategic planning [2]. In this perspective of computational thinking, computer science modeling techniques are essential in many aspects of modern-day research and in understanding things for all people who expect to live and work in a world where information is stored, accessed, and manipulated via computer software [2].

Wing [4] defined computational thinking as something that "involves solving problems, designing systems, and understanding human behavior, by drawing on the concepts fundamental to computer science". It includes [3]:

- A thought process, thus independent of technology.
- A specific type of problem-solving that entails distinct abilities (e.g., being able to design solutions that can be executed by a computer, human, or both).

However, Bocconi et al. [3] raised several issues and challenges that must be addressed for the effective integration of information technology in compulsory education, including *What are the core characteristics of computational thinking and its relationship with programming/coding in compulsory education?* Coding (programming) is regarded as a key 21st century skill: "Coding is the literacy of today and it helps practice 21st century skills such as problem-solving, modeling and analytical thinking" [3]. The authors of *European e-Skills Manifesto* [5] declared that "Skills like coding are the new literacy. Whether you want to be an engineer or a designer, a teacher, nurse or web entrepreneur, you'll need digital skills."

In this paper, we seek to contribute to the current debate on computational thinking with particular focus on the following.

### A. Conceptualization

In computer science, *conceptualization* is the first stage of the model-building process to arrive at a representation capable of addressing the relevant problem. A conceptual model is mainly formed upon concepts such as components of thinking. It can provide a framework for thinking that structures notions into patterns according to categories to provide a basis to represent internal thinking in an external form. Here, we use this modeling in the sense of patterned thinking [6] (e.g., object-oriented modeling), where pattern refers to recurring templates used by persons in the thinking process.

This paper promotes conceptual modeling that is based on the Heideggerian [7] notion of *thinging* as a framework for computational thinking. Heideggerian thinging is generalized as an abstract *thinging machine* (TM) [8-13].

### B. Core Concepts

As will be described in this paper, we propose five basic concepts to model computational thinking:

- The notion of *thing*;
- The notion of TM;
- Five flow operations of things: create, process, release, transfer, and receive; and
- Triggering.

*C. Programming/Coding*

A diagram can be coded, and the code and diagram approximate the conceptual form of the programmer behind both. A TM is expressed as a diagram that can be mapped to programming/coding in the same way as flowcharts. It is important to mention this property of the TM, even though it will not be explored in this paper.

To achieve a self-contained paper, Section II reviews the TM that was adopted in this paper and was used previously in several published papers, as mentioned previously. Section III presents examples of applying TM in computational thinking. Section IV applies the TM in an actual case study.

## II. THINGING MACHINE (TM)

Drawing on Deleuze and Guattari [14], who declared—admittedly from a different prospect—"All objects can be understood as machines," TM-based conceptual modeling utilizes an abstract *thinging machine* (hereafter, *machine*) with five stages of thinging, as shown diagrammatically in Fig. 1.

In philosophy, thinging refers to "defining a boundary around some portion of reality, separating it from everything else, and then labeling that portion of reality with a name" [15]. However, according to our understanding, thinging is when a thing manifests or unfolds itself in our conceptual space. An architect realizes the thing *house*, which in turn *things* (verb) [7]; that is, it presents its total *thingness*, which includes living space, shelter from natural elements, family symbol, etc. This issue will be explained later in this paper.

Our TM modifies Heidegger's [7] notion of thinging by applying it to the life cycle of a thing and not just to its ontological phase (producing). A thing things; in other words, a bridge is not a mere object; rather, it establishes itself in a conceptual realm as unified whole involving riverbanks, streams, and the landscapes. When representing it, we can view thinging as akin to an abstraction, but it differs in being expansive instead of being reductive in detail.

In the TM, we capture thinging as a dynamic machine of things that are created, processed, received, released, and transferred—the operations of Fig. 1. Heidegger [7] offered an example of thinging through the thing *jug*. When the clay is shaped into a jug, the jug manifests itself—in Heidegger's words—into "what stands forth." Its thingness conquests and entraps the void that holds and takes over its task of embracing and shielding the penetrating wine, thus connecting itself to a setting of vine, nature, etc. This conceptualization of the thing jug comes as a *reaction* to the physical formation of the clay. According to Heidegger, "We are *apprehending* it-so it seems-as a thing" [7] (italics added). The TM expands this thinging by conceptualizing the jug not only through its existence but also through its activities as a machine (an assemblage) that creates (e.g., certain shape of void), releases, transfers (e.g., air), receives, and processes other things. It is not only a thing that *things* but also a machine that *machines* (verb).

Heidegger [7] distinguished between objects and things: "The handmade jug can be a thing, while the industrially made can of Coke remains an object" [16]. The industrially made can of Coke has minimal thinging and maximal abstracting (see later discussion). Note that this does not apply to other industrial devices that are not cut off from their "roots." The thermostat, for example, is an industrial product that manifests itself in its environment, as will be represented later in this paper. For Heidegger [7], things have unique "thingy Qualities" [16] that are related to reality and therefore are not typically found in industrially generated objects. According to Heidegger [7], a thing is self-sustained, self-supporting, or independent—something that stands on its own. The condition of being self-supporting transpires by means of *producing* the thing. According to Heidegger [7], to understand the thingness of a thing, one needs to reflect on how thinging expresses the way a "thing *things*" (i.e., "gathering" or tying together its constituents into a whole). According to Thomas et al. [17], Heidegger's view can however be seen as a tentative way of examining the nature of entities, a way that can make sense. An artefact that is manufactured instrumentally, without social objectives or considering material/spatial agency, may have different qualities than a space or artefact produced under the opposite circumstances.

The TM handles things and is itself a thing that is handled by other machines. The stages in the machine can be briefly described as follows:

**Arrive:** A thing flows to a new machine (e.g., packets arrive at a buffer in a router).

**Accept**: A thing enters a machine; for simplification purposes, we assume that all arriving things are accepted; hence, we can combine **Arrive** and **Accept** into the **Receive** stage.

**Release**: A thing is marked as ready to be transferred outside the machine (e.g., in an airport, passengers wait to board after passport clearance).

**Process** (change): A thing changes its form but not its identity (e.g., a number changes from binary to hexadecimal).

**Create**: A new thing is born in a machine (e.g., a logic deduction system deduces a conclusion).

**Transfer**: A thing is inputted or outputted in/out of a machine.

A TM also utilizes the notion of *triggering*. Triggering is the activation of a flow, denoted in TM diagrams by a dashed arrow. It represents a dependency among flows and parts of flows. A flow is said to be triggered if it is created or activated by another flow (e.g., a flow of electricity triggers a flow of heat) or activated by another point in the flow. Triggering can also be used to initiate events such as starting up a machine (e.g., remote signal to turn on). Multiple machines can interact by triggering events related to other machines in those machines' stages.

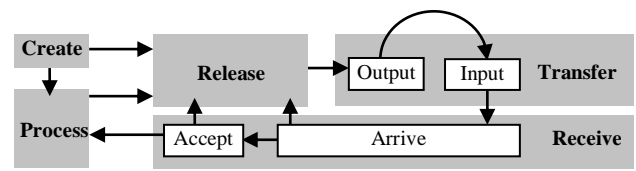

Fig. 1. Thinging Machine.

## III. EXAMPLE

According to Riley and Hunt [2] in their book *Computational Thinking for the Modern Problem Solver*, an abstraction is anything that allows us to concentrate on important characteristics while deemphasizing less important, perhaps distracting, details. Abstraction is a core concept in computational thinking in addition to such notions as algorithmic thinking, decomposition, and generalization [3]. Riley and Hunt [2] stated that programmers are really a kind of problem solver and that computer programmers are arguably the most important of all modern problem solvers. The best way to characterize computational thinking is through the way computer scientists think, as well as the manner in which computer scientists think for the rest of us. As a digital camera uses a handful of focus points, computer scientists learn to focus on the most important issues through abstraction [2].

The notion of abstraction goes all the way back to Plato, who proposed to distinguish abstract ideas as ideal entities that capture the essence of things. They are abstraction, that is, ideas that do not exist in the world. We can note two basic aspects of abstraction:

- Not being in reality,
- Being reductive in details

Abstraction is an important way of thinking, nevertheless,

We claim that thinging is also a valuable process as a fundamental skill for everybody in computational thinking.

Thinging takes a holistic view by, in contrast to abstraction, being *expansive* in detail, as shown in Fig. 2. Thinging is an abstraction-like process that deemphasizes reduction and hence facilitates seeing the "bigger picture." Note that thinging and abstraction can be performed at several levels of expansion and in reduction of details. Fig. 3 illustrates the nature of thinging as an inverse of realization in reality.

Note the reductive nature of *object*-oriented modeling (e.g., UML) in the following example. As shown in Fig. 4, Riley and Hunt [2] *abstractly* described the thermostat, which involves a class diagram rectangle consisting of three parts diagrammed in three compartments. The middle compartment lists attributes of the thermostat. The operations in a class diagram are listed in the bottom compartment, where operations are abstract references to the behavior of the object. The following model presents an alternative conceptualization of the thermostat.

### A. Static TM of the Thermostat

The thermostat can be represented as in Fig. 5. In line with the previous discussion on the thermostat, its thingness includes Switch (1), Fan (2), and Temperature (3). The switch includes three signals, COOL (4), OFF (5), and HEAT (6), which flow to change the State (7) of the cooling/heating machine (8). Similarly, signals set the temperature (9) and change the state of the fan (10).

### B. Behavior of the Thermostat

Behavior in a TM is represented by *events*. An event is a thing that can be created, processed, released, transferred, and received. It is also a machine that consists of (at least) three submachines: region, time, and the event itself. As a side note, we may conceptualize the TMs as fourfold—that is, consisting of space, time, event, and things.

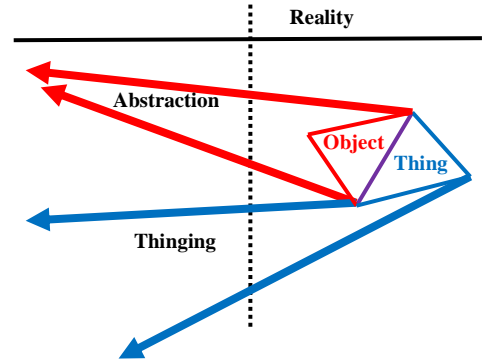

Fig. 2. Thinging is an Expansive Reverse of Realization in Reality.

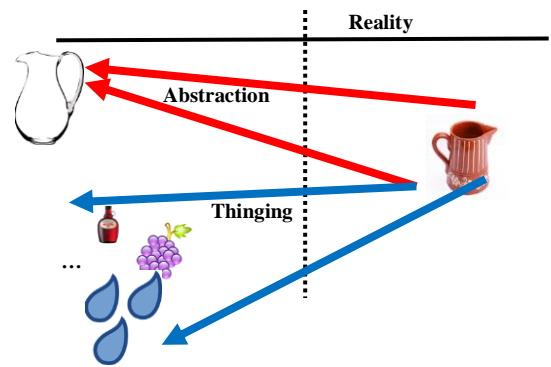

Fig. 3. The Thing Jug things through its Total Thingness.

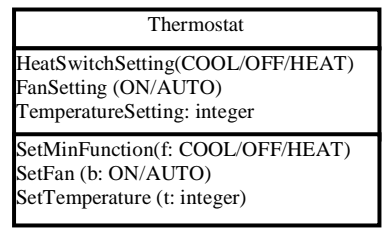 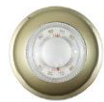

Fig. 4. Description of the Class Temperature (Adapted from [2]).

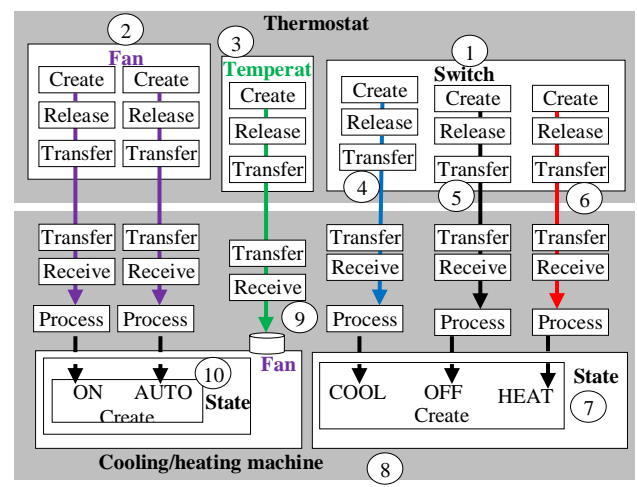

Fig. 5. The TM Representation of the Thermostat.

Consider the event *The switch turns OFF* (see Fig. 6). It includes the event itself (Circle 1 in Fig. 6), the region of programmers the things currently being dealt with in the event (2), and the time machine (3). The region is a subgraph of the static representation diagram of Fig. 5. For simplicity's sake, we will represent an event by its region only.

Accordingly, we can identify four basic events in the static description of Fig. 5, as shown in Fig. 7:

- Event 1 ($E_1$): The switch is COOL.
- Event 2 ($E_2$): The switch is OFF.
- Event 3 ($E_3$): The switch is HEAT.
- Event 4 ($E_4$): The temperature is SET.
- Event 5 ($E_5$): The fan is ON.
- Event 6 ($E_6$): The fan is AUTO.

These events can be written as statements of any programming language.

*C. Control of the Thermostat*

A possible events chronology is shown in Fig. 8, which represents the permitted sequence of events. For example, switching directly from COOL to HEAT and vice versa without first turning the cool/heat machine OFF is not permitted. These sequences are shown in Fig. 9 (a-e) as follows:

*1)* The cool/heat machine is OFF,

   *a)* Select {COOL or HEAT}, then fan {ON fan, set the temperature}.

   *b)* Select HEAT {select the state of the fan, set the temperature}.

*2)* The cool/heat machine is on {COOL or HEAT}, and the fan is {ON or AUTO}, switch fan to {ON or AUTO}.

*3)* The cool/heat machine is on {COOL or HEAT}, set the cool/heat machine OFF.

*4)* The cool/heat machine is on {COOL or HEAT}, set the temperature.

*5)* The cool/heat machine is OFF, switch fan to {ON or AUTO}.

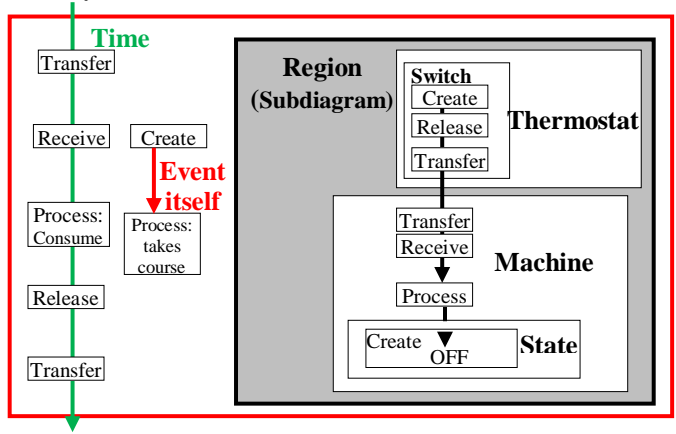

Fig. 6. He Event: the Switch Turns OFF.

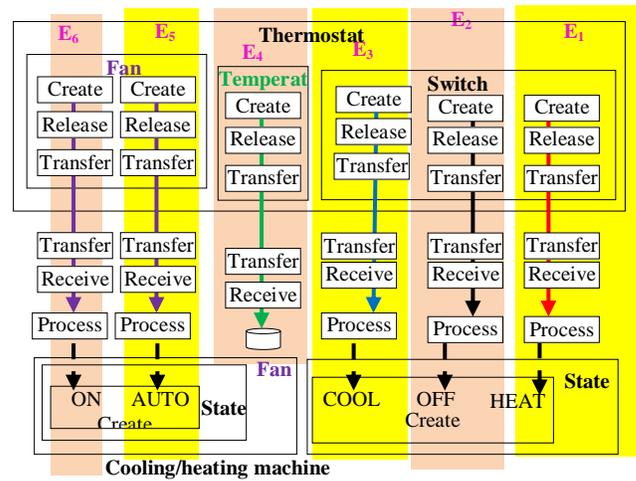

Fig. 7. The Events of the Thermostat.

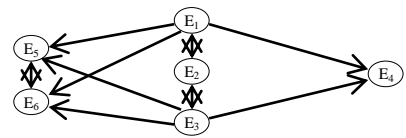

Fig. 8. Chronology of Events.

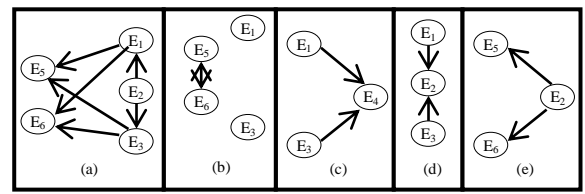

Fig. 9. Permitted Sequence of Control Operations.

*D. Mapping to Class Notations*

Selecting the events is a design decision. TM representation shows that Riley and Hunt [2] declared only three events (Fig. 10):

- Event 1 ($E_1$): The switch is COOL/OFF/HEAT.
- Event 2 ($E_2$): The fan is OFF/AUTO.
- Event 3 ($E_3$): The temperature is set.

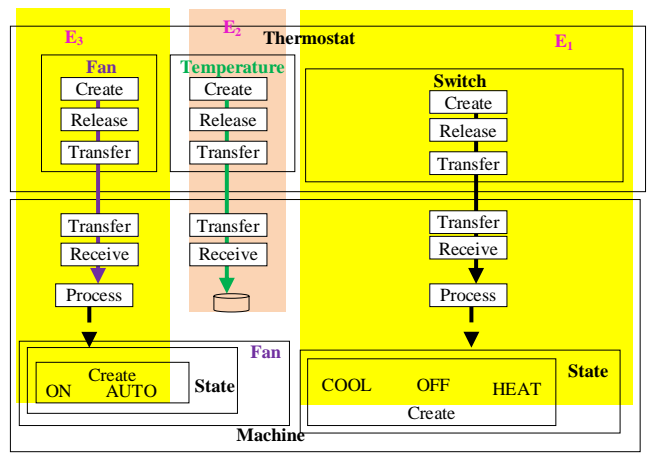

Fig. 10. The Events of the Thermostat.

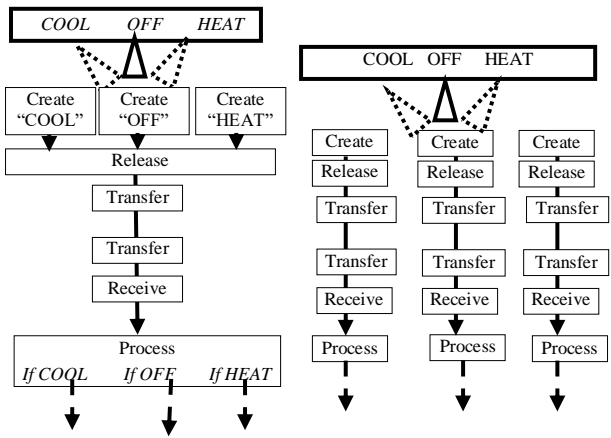

Fig. 11. The Switch Representation in the 3-Events (Left) and 6-Events (Right) Designs of the Thermostat.

Fig. 11 contrasts the switch representation in the 3 and 6 designs.

The class notation given by Riley and Hunt [2] can be viewed as mere names for data items and methods (processes) that can be mapped to the TM, as shown in Fig. 12. Thus, we can produce the class description from the TM representation.

The important point is that the object-oriented thinking style, the class description, is produced before describing the methods, whereas in the TM, the TM machines are developed right from the beginning of the analysis. Designing the thermostat in terms of three events is the result of this object orientation, which captures the three events because it does not see all the possibilities of design.

Consider the 3-events and 6-events designs. The 3-events uses one wire between the thermostat and the cool/heat machine, whereas the 6-events design uses three. Each implementation has its merits. The 3-events design is cheaper, and the 6-events is more reliable. For example, in the 6-events design, if heating does not work, the cooling feature will still work when the link to the cool/heat machine is cut. The point here is that the object-orientation, as discussed by Riley and Hunt [2], does not seem to be aware of available alternative designs. This is an important observation in the context of thinking. According to Do and Gross [18], in design, "Drawing is intimately bound with thinking."

## IV. CASE STUDY

The thermostat's TM modeling is a small artificial example of problem-solving by describing it conceptually. Our case study involves a large real problem: how to model a help desk in a government ministry. In its actual environment (the workplace of the second author), the maintenance process starts when a user contacts the IT department for help. The department calls such a process the help desk process. It is a problematic system that involves implicit contacts and interactions in the alignment between IT and business [19].

In this case study, the IT department solved the help desk problems using an ad-hoc technique that involves thinking of it as a semi-automated system that is built piece by piece over several years. There is no current documentation, even though the manager of the help desk drew flowcharts that show the full description of the processes behind how the help desk works for different tasks, as shown in Fig. 13. In projecting this system on Heidegger's jug, in such an approach, this can be viewed as failure to give thought to "what the jug holds and how it holds".

Help desk operations are causing many types of managerial, supervision, technical, and legal problems. A possible solution is a holistic approach that involves all related elements in the help desk system. It is a system that exists in reality and needs a better understanding of its thinging. It is *misthinged* or, in Heideggerian language, a broken tool that marks the annihilation of the "equipmental thing" (IT help desk), in that helping cannot be gathered around it.

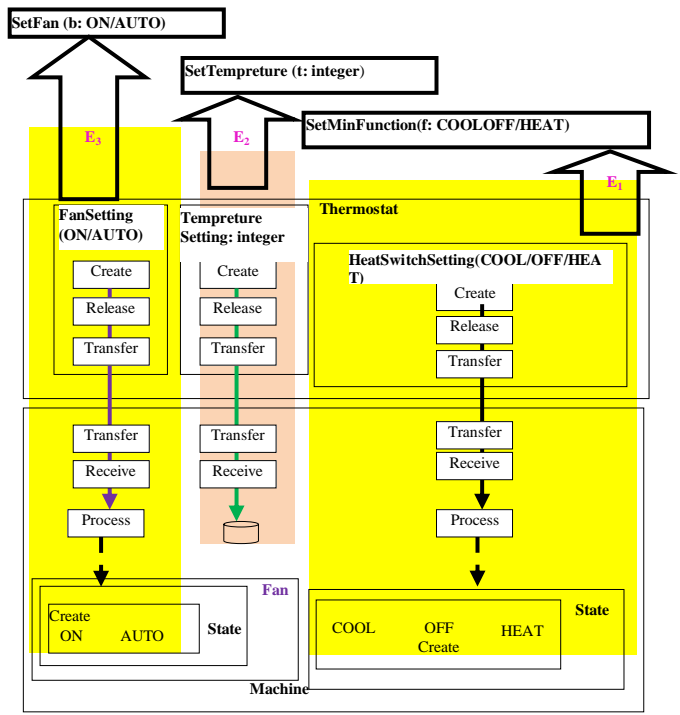

Fig. 12. TM and Class Entries.

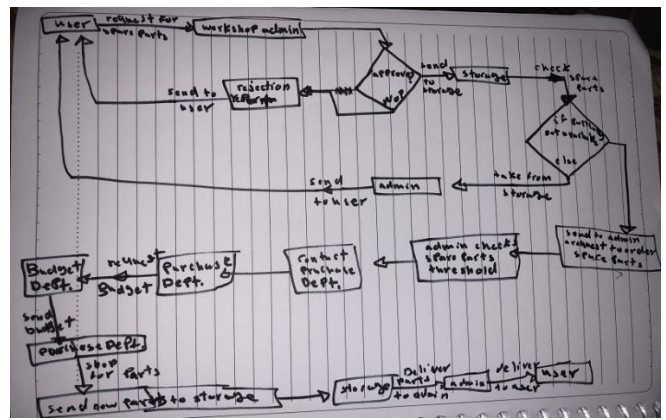

Fig. 13. Sample Current Documentation.

Accordingly, we consider the question: "How does the IT help desk operate?" We conceptualize it as a TM that creates, processes, releases, transfers, and receives things. The helping system includes things that are machines and machines that are things unfolding an integrated wholeness that is itself part of the ministry's machinery. We focus next on thinging the IT help desk.

*A. Static Model*

Accordingly, we model the help desk system, as shown in Fig. 14. In the figure, the user sends a request to the secretary of the workshop (Circle 1). The request is checked to decide whether it is for repair (A) or for spare parts (B).

*B. Request for Repair*

The repair request flows to the workshop administrator (2), where it is processed to do the following:

*1)* Selecting a specific technician for this request: To accomplish that, the list of technicians is processed (4) to generate the name of a technician (5).

*2)* Creating a task (ticket): Additionally, the administrator creates a new task form (6) that includes the request description (7) and the technician's name (8).

The task then flows (9) to the technician, who later examines the task to decide on the following:

*1)* Given that it is possible to call the user and solve the problem by phone (10), the technician places a phone call (11) to the user and guide the user step by step to solve the problem through the phone (12).

*2)* The technician is required to go to the user's workplace (13) to solve the problem by him-/herself (14). The technician moves from the workshop to the user's location (15). The user brings the computer to the technician to work on it and repair it (16).

After processing the computer (17), the technician has one of the two following outcomes:

*1)* The computer is not repaired (18), and the technician takes it back to the workshop. There, it is fixed (19), and the workshop admin (20) transfers the fixed computer back to the user (21).

*2)* The computer is repaired (22) and transferred back to the user (23 and 24).

Both previous outcomes lead to (25), where the user gets the computer and processes it to see whether it is repaired:

*1)* The computer works fine (26); as a result, the user creates a report (27) to close the request and sends this report to the workshop admin (28).

*2)* The computer repair is not satisfactory (29), and the user creates a follow-up request (30) for repair and sends it to the secretary (A).

Request for spare parts

The spare parts request flows to the inventory department (31), where it is processed (32) to extract the quantity of current spare parts in the inventory (33) and to transfer it to a program that checks this quantity of spare parts (34):

*1)* If the number is zero, the number of the pending requests would be incremented by one (35). Moreover, the request would be released (36) and added to a queue of pending requests (37).

*2)* If the number is greater than zero, the request is processed again (38 and 39) to extract the requested quantity of spare parts (40).

Note that we renovated an existing system and did not design the best model for this application. For example, it is possible to define the minimum value of inventory instead of permitting it to reach zero. Thus, our thinging of the system is tailored to the existing requirements.

Both the numbers of the requested items (41) and current quantity (42) are transferred to a program that calculates the available quantity (43) that can be delivered to the requester. A simple formula calculates what is called *remaining quantity* as follows:

*Remaining Quantity = Current Quantity – Requested Quantity (44)*

Accordingly, two possibilities arise:

*1)* The remaining quantity is greater than or is equal to zero (45); in other words, the full requested quantity can be provided to the user. In that case, the request is released (46) and transferred to the storage, where it is received and processed (47) and the stored spare parts are sent to the requester (48).

*2)* The remaining quantity is less than zero (49); as a result, a new quantity called *pending* is created and calculated as the following:

*Pending = Requested Quantity – Current Quantity*

Accordingly, a new request that specifies the quantity that is currently in the possession of the inventory department is created (50) and forwarded to the storage, and then steps (46-48) are repeated. Also, a new request that specifies the number of pending quantities is created and considered as a new request (51).

In parallel, according to a certain schedule (52), the list of pending requests is processed, and each request (the loop is specified in the dynamic TM model) is taken out and processed to create a pending request (53) that, in turn, is processed, thus leading to the creation of an ordered quantity (54). The ordered quantity is added to the total number of ordered items (55). Later, the total number of ordered items (56), along with the current quantity (57), flows to a committee for examination, and the evaluation of the need for new spare parts is processed (58). Hence, a decision is created (59) and processed for making orders (60), which flow to the workshop admin (61).

In the workshop admin, the orders are processed to (62) create orders to the suppliers (63) and transfer these orders to the purchase department (64). There, each order is processed (65) and put on hold while waiting to assign a budget (66). A request for a budget is created (67) by the purchase department and is transferred to the budget department (68). The budget department processes the budget request, (69) approves it, and then sends the approval to the purchase department (70). In the purchase department (71), the approval is processed, thus leading to placing an order to the supplier (72).

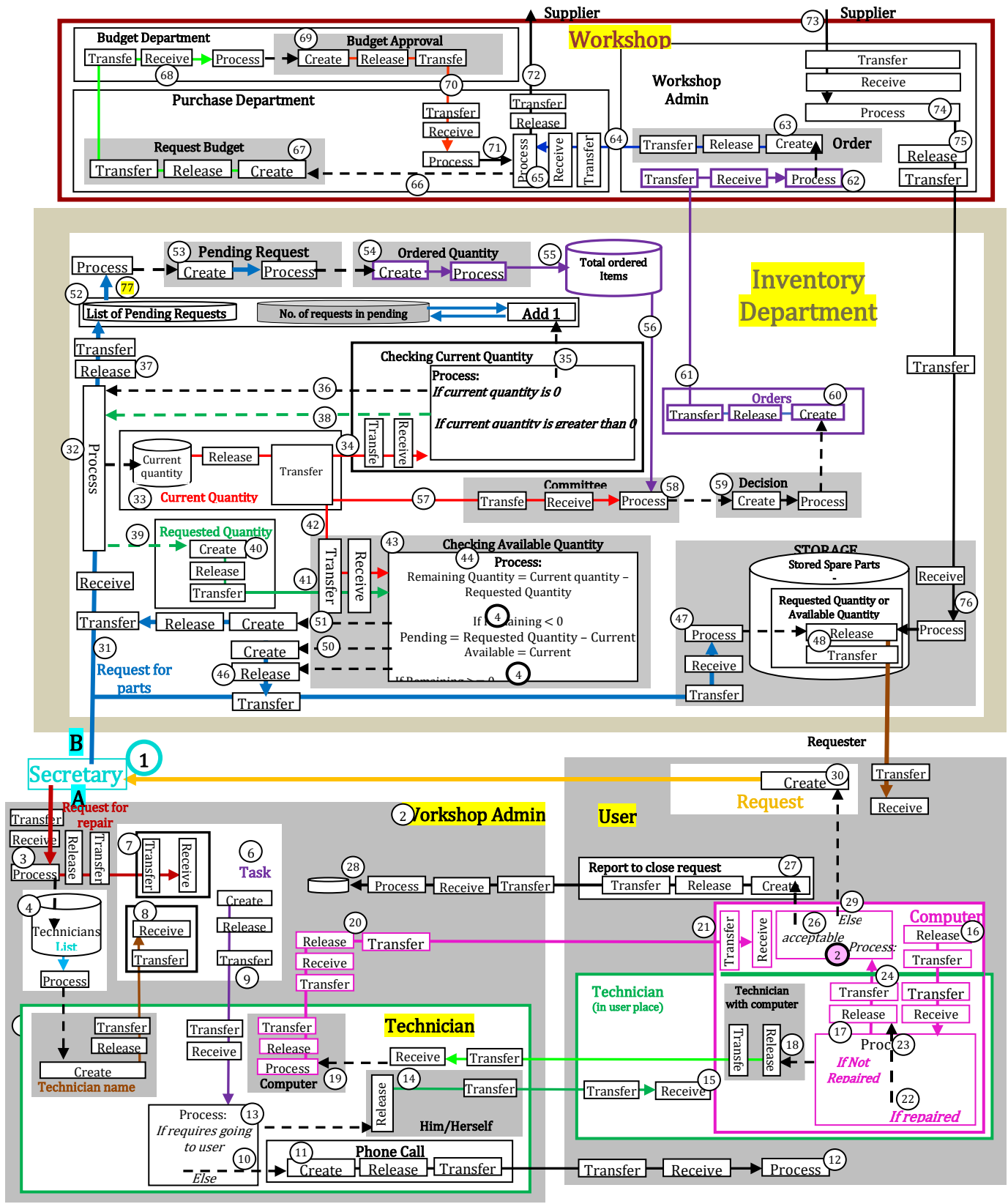

Fig. 14. The TM Representation of the IT Department Help Desk System.

(IJACSA) International Journal of Advanced Computer Science and Applications,
Vol. 10, No. 2, 2019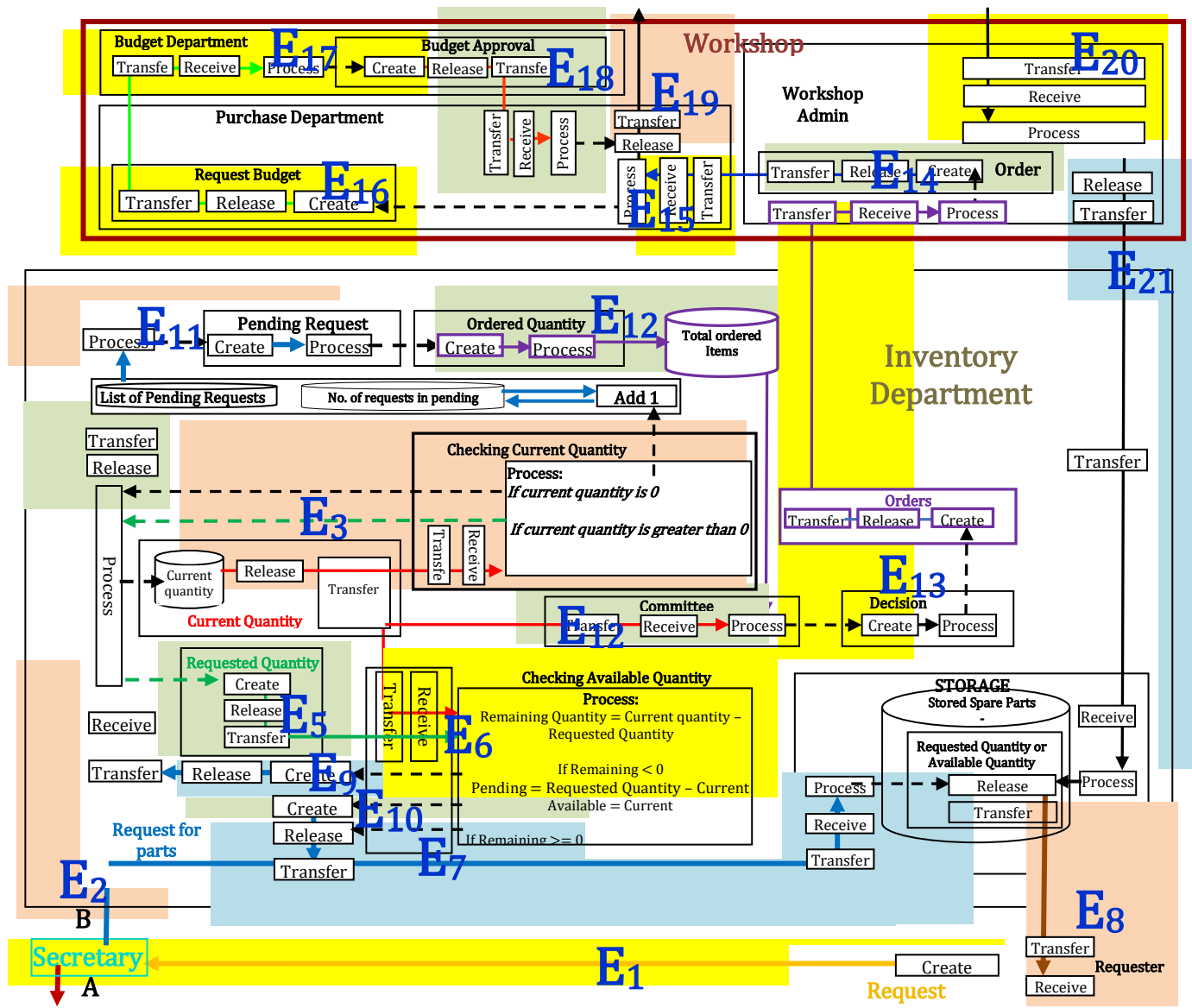

Fig. 15. Events of the TM Representation of the IT Department Help Desk System (Partial).

*C. Behavior Model*

As mentioned previously in the thermostat example, behavior in a TM is represented by *events*. Accordingly, we can identify the following events in the static description of Fig. 14, as shown in Fig. 15. To save space, we identify only the upper part of Fig. 14 (requesting parts):

- Event 1 ($E_1$): The secretary receives a request for purchasing spare parts.

- Event 2 ($E_2$): The inventory department receives and processes the request.

- Event 3 ($E_3$): The current quantity is retrieved and processed.

- Event 4 ($E_4$): If the current quantity is 0, add the request to the pending requests list and update the number of pending requests.

- Event 5 ($E_5$): If the current quantity is greater than 0, extract the requested quantity.

- Event 6 ($E_6$): Find Remaining (Quantity = Current quantity – Requested Quantity) and process it.

- Event 7 ($E_7$): Given that Remaining $>= 0$, retrieve the requested items from the Storage.

- Event 8 ($E_8$): Send the requested items to the requester.

- Event 9 ($E_9$): If Remaining $< 0$, calculate Pending = Requested (Quantity–Current), create a request for pending items, and add the request to the list of pending requests.

- Event 10 ($E_{10}$): If Remaining $< 0$, calculate Available = Current and retrieve the requested items from the storage.

- Event 11 ($E_{11}$): Retrieve the pending requests and extract the requested quantities.
- Event 12 ($E_{12}$): Both requested pending quantities and current quantities are sent to the ordering committee.
- Event 13 ($E_{13}$): The committee creates orders and sends them to the workshop.
- Event 14 ($E_{14}$): Orders are received by the workshop and orders to the supplier are created.
- Event 15 ($E_{15}$): The purchase department receives orders for the supplier.
- Event 16 ($E_{16}$): A request for budget is created.
- Event 17 ($E_{17}$): The request for budget flows to the budget department.
- Event 18 ($E_{18}$): The budget is approved.
- Event 19 ($E_{19}$): Orders for the supplier are sent.
- Event 20 ($E_{20}$): Ordered items are received from the supplier.
- Event 21 ($E_{21}$): Items as sent to the storage.

Fig. 16 shows the chronology of these events.

### D. Control

Control can be superimposed onto the events of the TM system. In the case study, suppose that we want to declare the following warning messages related to the management of the system:

*1)* If the time to order from the supplier in the workshop exceeds t1, then create a warning message.

*2)* If the time to deliver items received from the supplier to the requester exceeds t1, then create a warning message.

Fig. 17 shows the declaration of these rules over the chronology of events. In Fig. 18, when the workshop receives an order, the time of the order arrival is created. This time is processed repeatedly. If the time exceeds t1—the time period since the receiving of the order—then a warning is created. A similar process is followed for the second rule.

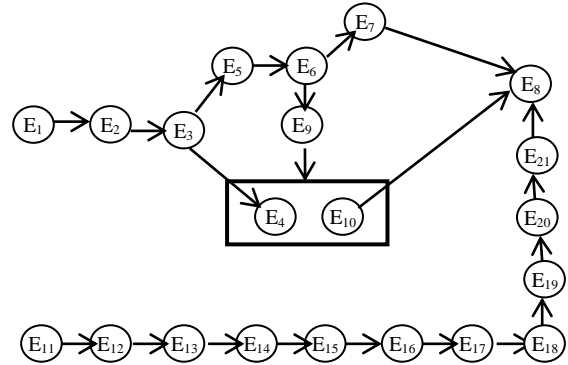

Fig. 16. The Chronology of Events of the Case Study.

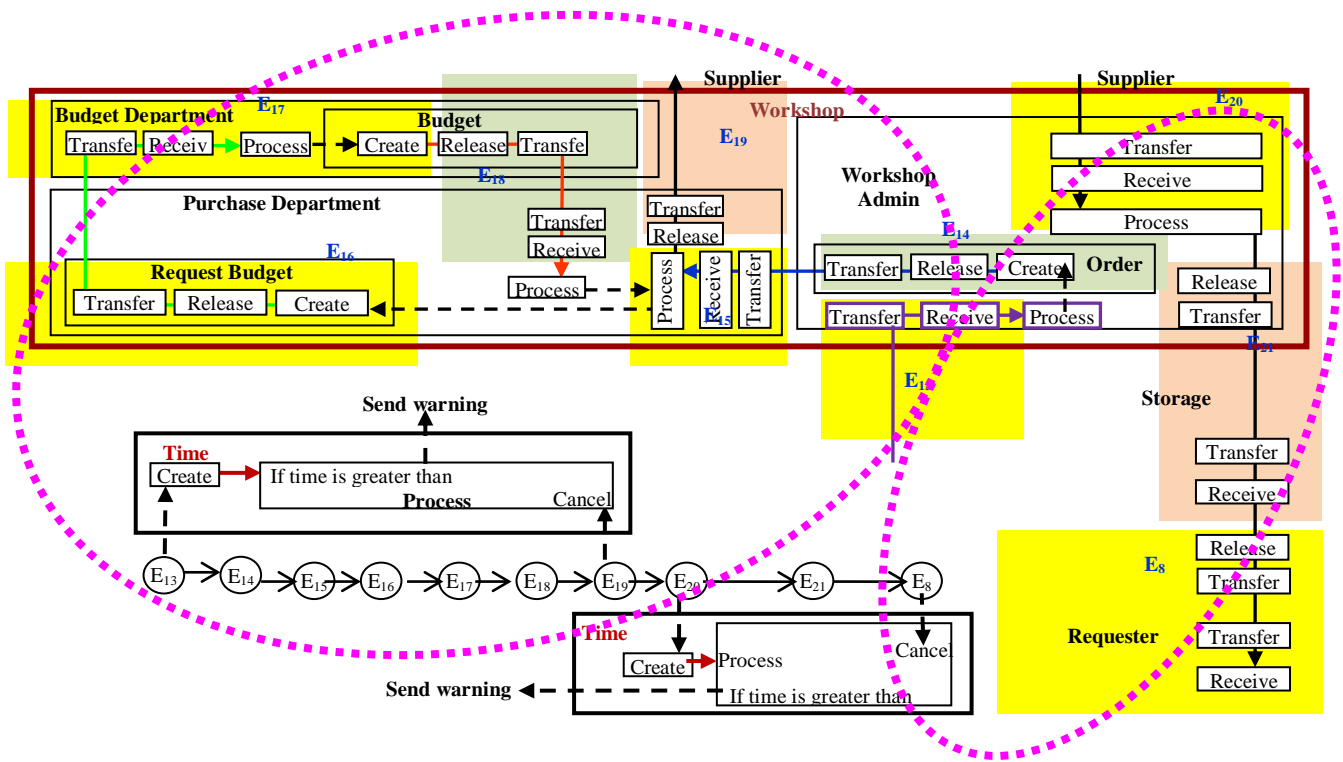

Fig. 17. Examples of Control in the Case Study.

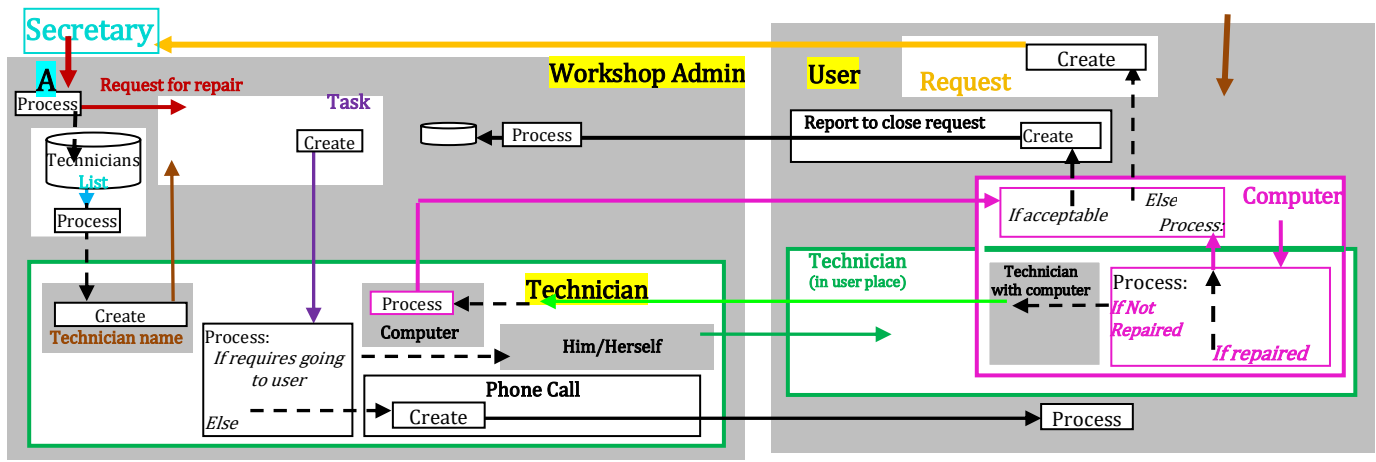

Fig. 18. Simplification of the TM Representation of the IT Department Help Desk System by Removing the Stages Transfer, Release, and Receive.

## V. Conclusion

We proposed using a new modeling technique, TM, as a foundation in computational thinking. According to the TM approach, a person's "thought machine" forms a train of thought that excludes other modes such as procedural and object-oriented modes of thinking. The paper emphasizes this thinking style as a unifying method that could have diverse applications. The TM is an underlying tool for expressing the unified totality of a system's things and machines analogous to carpeting techniques where a ground fabric beneath the design binds pieces and sews the patterns of fabric.

To substantiate our claim, we contrast the TM side by side with diagrams of other approaches (e.g., the thermostat). Although we provided comprehensive evidence of our claim, its inaccuracy or its partial value needs efforts beyond a single researcher. However, the thermostat example and the case study seem to point to some merits that deserve more development.

Fig. 14 of the case study may raise the issue of the TM diagram's complexity. The TM model can be specified at various levels of granularity. For example, Fig. 18 is a simplified version of the lower part of Fig. 14. The stages transfer, release, and receive are deleted under the assumption that the direction of the flow arrow is sufficient to represent them.